\documentclass[11pt]{article}

\usepackage{epsfig}
\usepackage{cite}
\usepackage{amsmath,amsfonts,amssymb}
\usepackage{mathbbol}

\numberwithin{equation}{section}

\newcommand{\N}{\ensuremath{{\cal N}}}

\newcommand{\half}{\ensuremath{\frac{1}{2}}}

\newcommand{\quarter}{\ensuremath{\frac{1}{4}}}

\newcommand{\be}{\begin{equation}}
\newcommand{\ee}{\end{equation}}

\newcommand{\ba}{\begin{eqnarray}}
\newcommand{\ea}{\end{eqnarray}}

\newcommand{\ns}{\normalsize}

\newcommand{\gsim}{\raise.3ex\hbox{$>$\kern-.75em\lower1ex\hbox{$\sim$}}}
\newcommand{\lsim}{\raise.3ex\hbox{$<$\kern-.75em\lower1ex\hbox{$\sim$}}}

\begin{document}

\begin{titlepage}

\title{
   \hfill{\ns hep-th/0510269\\}
   \vskip 2cm
   {\Large\bf The effects of inhomogeneities on the cosmology of type IIB conifold transitions.}
\\[0.5cm]}
   \setcounter{footnote}{0}
\author{
{\ns \large 
  \setcounter{footnote}{3}
  Eran Palti$^1$ \footnote{email: e.palti@sussex.ac.uk}, 
  Paul Saffin$^{1,2}$\footnote{email: paul.saffin@nottingham.ac.uk}, 
  Jon Urrestilla$^{1,3}$ \footnote{email: jon@cosmos.phy.tufts.edu} }
\\[0.5cm]
   $^1${\it\ns Department of Physics and Astronomy, University of Sussex}\\
   {\it\ns Falmer, Brighton BN1 9QJ, UK} \\[0.2em] 
   $^2${\it\ns School of Physics and Astronomy, University of Nottingham,}\\
   {\it\ns University Park, Nottingham, NG7 2RD}\\[0.2em]
   $^3${\it\ns Institute of Cosmology, Department of Physics and
   Astronomy}\\
   {\it\ns Tufts University, Medford, MA 02155,USA}
}

\date{}

\maketitle

\begin{abstract}

\noindent
In this paper we examine the evolution of the effective field theory describing
a conifold transition in type IIB string theory. Previous studies have considered such
dynamics starting from the cosmological approximation of
  homogeneous fields ,
here we include the effects of inhomogeneities by using a real-time lattice field
theory simulation. By including spatial variations we are able to simulate the effect
of currents and the gauge fields which they source. We identify two different regimes
where the inhomogeneities have opposite effects, one where they aid the system to
complete the conifold transition and another where they hinder
it. The existence of quantized fluxes in related systems has lead to the speculation
that (unstable) string solutions could exist, using our simulations we give strong
evidence that these string-like defects do not form.
\end{abstract}

\thispagestyle{empty}

\end{titlepage}

\section{Introduction}
\label{sec:introduction}

Calabi-Yau manifolds have long been known to exhibit topology changing transitions \cite{Candelas:1988di,Candelas:1989ug,Greene:1996cy}.  
The use of effective field theories to model these transitions was initiated in \cite{Strominger:1995cz,Greene:1995hu,Greene:1996dh} 
where it was shown that, by including non-perturbative states (associated with branes wrapping degenerating cycles) which become light 
near the transition points, there are smooth four dimensional descriptions of the transitions.  
This prompted studies of the transitions as explicit
time dependent phenomena. The case of flop transitions, whereby one $S^2$ is shrunk and replaced by a topologically
distinct $S^2$, was studied in \cite{Brandle:2002fa,Jarv:2003qx}.
This was followed by a study of the more drastic topology changing transition of the
conifold \cite{Mohaupt:2004pq,Mohaupt:2004pr,Lukas:2004du,Mohaupt:2005pa}, where an $S^2$ is replaced by, or replaces, an $S^3$.
An important aspect of the transitions which emerged from these studies is the issue of moduli trapping.
The cycles that are wrapped correspond to moduli of the Calabi-Yau and  
it was shown that the points in moduli space where the transitions occur, the discriminant locus, are natural attractors and 
will typically trap the moduli at the transition points. Cosmological effects were found to be particularly important for this to
occur as the moduli perform Hubble damped oscillations about the transition point. 
 In the case of flop transitions moduli trapping meant the transition could not 
be completed, but in the case of the conifold this was crucial for the transition to occur. 
In \cite{Lukas:2004du} it was shown that there is a possibility for the transition to occur if the moduli were trapped efficiently 
enough and the vacuum expectation values of the light states were large enough. 
Once the transition was completed, however, the vacuum expectation values (vevs) of the new light states remained as flat directions. These
could be thought of as the new moduli arising from the change in the Hodge numbers of the Calabi-Yau space. 

Another issue, first raised in \cite{Greene:1996dh}, is
  that there is flux quantization in the system, leading the
  authors to predict the possibility of string formation as a mechanism
  for confinement. However, it was known that the existence of quantized fluxes
  is not enough to ensure the existence of stable string
  configurations, especially when global and local symmetries are
  involved (as is the case here). The classic example is that
  of semi-local strings, where stability of the vortices
depends on the parameters of the model \cite{Hindmarsh:1991jq,Achucarro:1999it}.
 It was subsequently shown in \cite{Achucarro:1998er} that 
the infinite axi-symmetric ``semilocal'' strings 
of the conifold transition are unstable for {\it all} values
  of the parameters
 as they can always lower their energy through an expansion of the
  core. Different possible mechanisms to stabilize these strings were
  studied in a series of papers \cite{Achucarro:2001ii,Pickles:2002ym,Achucarro:2002jg}, but the only possibility to
  find stable infinite vortices was for supersymmetry to be softly broken by adding
  mass terms to the scalars\footnote{There might be a
  possibility of having confinement for semilocal-like string solutions for finite strings \cite{Evlampiev:2003ji}.}. The issue of whether the strings
form at all and on what time scales they decay, however, was not studied.

In this paper we aim to address the issues of moduli trapping and string formation from a full three dimensional point of view,
allowing spatial variations of the fields.
The cosmological models \cite{Brandle:2002fa,Jarv:2003qx,Mohaupt:2004pq,Mohaupt:2004pr,Lukas:2004du,Mohaupt:2005pa} were, for simplicity, 
treated as effectively one dimensional with the fields taken as 
spatially homogeneous. This has the effect that no currents are formed, which in turn means that gauge fields will not
be excited, so any dynamics involving these currents and gauge fields has not been studied.
Although this is a pragmatic approximation, in many circumstances we know that interesting physics can emerge
from finite correlation lengths, such as topological defects \cite{vilenkin:1994}.
By performing full three dimensional numerical simulations of the transition we will show that inhomogeneities in the fields have
a complicated effect on the issue of moduli trapping and may facilitate or hinder the process depending on their amplitude.
Cosmic string formation in conifold transitions would have provided an interesting possibility for experimental signature, however,
although gauge fields are induced through spatial currents, the 
``semilocal''  strings discussed above
do not form.  
The paper will be set out as follows.     
In section \ref{sec:thetheory} we will introduce the fields and the action we will be working with. 
We will then discuss the issue of moduli trapping in more detail summarising 
the findings in \cite{Brandle:2002fa,Jarv:2003qx,Mohaupt:2004pq,Mohaupt:2004pr,Lukas:2004du,Mohaupt:2005pa} and introduce the alternative
 idea of quantum moduli trapping \cite{Kofman:2004yc}. In section \ref{sec:simulations} we will introduce the details of 
the simulations that were performed and discuss the results. We conclude the paper in section \ref{sec:conclusions}.

\section{The effective theory}
\label{sec:thetheory}

In this section we will consider the effective theory that models the transition. In section \ref{sec:action} we will 
introduce the action and the fields involved. We will then discuss the idea of moduli trapping in section \ref{sec:modulitrapping}

\subsection{The action}
\label{sec:action}

The particular case we will be considering is a conifold transition in type IIB string theory. 
We will briefly summarise the important elements below but for a 
more complete summary of the geometry and physics involved see \cite{Lukas:2004du} and references therein. 
We will consider the case of two degenerating three-cycles\footnote{Two is the smallest number of cycles that must simultaneously degenerate in 
order for the transition to occur between two Calabi-Yau manifolds \cite{Candelas:1989js,Greene:1996dh}.}
with an homology relation between them so that they can both be described in terms of one complex structure modulus.
Such a modulus can be defined by integrating the unique holomorphic three form on the complex structure moduli space $\Omega$ over one of the cycles

\be
Z \equiv \half \left(M + iN \right) \equiv \int_{\cal A} \Omega 
\ee
where ${\cal A}$ denotes one of the degenerating cycles and $M$ and $N$ are defined by the above relation.
The conifold point then corresponds to $|Z|=0$. The non-perturbative states, corresponding to branes wrapping the cycle, 
that become light near the conifold point can be described by the scalar components of $\N=2$ hypermultiplets. For the case
of two degenerating cycles there are two such hypermultiplets and we denote their scalar components by $q^{au}$ where $a,b,..=1,2$ denote
the hypermultiplet and $u,v,..=0,1,2,3$ run over the hypermultiplet components.   
The theory we will be working with is a gauged ${\cal N}=2$ super Yang-Mills with the action
\ba
S &=& \int -\half \partial_{\mu} M \partial^{\mu} M - \half \partial_{\mu} N \partial^{\mu} N  \nonumber \\
 &-& \left( \partial^{\mu} q^{1u} + e A^{\mu} t^u_{\;v} q^{1v}
\right)\left( \partial_{\mu} q^{1u} + e A_{\mu} t^u_{\;w} q^{1w} \right) \nonumber \\
 &-& \left( \partial^{\mu} q^{2u} - e A^{\mu} t^u_{\;v} q^{2v}
\right)\left( \partial_{\mu} q^{2u} - e A_{\mu} t^u_{\;w} q^{2w} \right) \nonumber \\
 &-& \quarter F_{\mu\nu} F^{\mu\nu} \nonumber \\
 &-& \left( M^2 + N^2 \right) \left( q^{1v}q^{1v} + q^{2v}q^{2v} \right) \nonumber \\
 &-& 2\left( \quarter q^{1v}q^{1w}q^{1v}q^{1w} + \quarter q^{2v}q^{2w}q^{2v}q^{2w} - q^{1v}q^{1w}q^{2v}q^{2w}  \right. \nonumber \\
 && \;\; \left. + \half q^{1v}q^{1v}q^{2w}q^{2w} - q^{1v}q^{1w}q^{2r}q^{2t}t_{wt}t_{vr} \right) \label{action}
\ea
where
\ba
& &t^{1}_{\;2} = -t^{\;1}_{2} = 1 = t^{3}_{\;4} = -t^{\;3}_{4} \\
& &F^{\mu\nu} = \partial^{\mu} A^{\mu} - \partial^{\nu} A^{\mu}
\ea 
The charge $e$ is unity in our units but we leave it in so that we can study the case where the scalars decouple from the gauge fields by setting it to zero.
We have set the four-dimensional Plack constant to unity thereby fixing our units.
The gauge field $A^{\mu}$ completes the bosonic components of the $\N=2$ vector multiplet that contains the scalars $M$ and $N$.
The action (\ref{action}) is not a supergravity and so cannot be used for cosmological models. However, we will now justify that
the action can be supplemented by Hubble friction in the equations of motion such that it will capture all the essential features
of the physics involved. The above action
differs from a supergravity in three ways. The first is that the Ricci scalar is missing. This term would lead to Hubble friction terms $ \sim 3H\dot{q}$ in the equations of motion, with $H$ given by the Friedman equation. We will incluide those Hubble friction terms in the numerical equations of motion.
The second is that the metric on the moduli space of $Z$ is taken to be flat while in the
full supergravity it will not be and would have to be calculated from the complex structure prepotential. However, in this paper
we will be mainly considering the evolution of the light states $q^{au}$ rather than the evolution of $Z$ and will fix $Z$ to be 
at the conifold point. In that case the metric on the space does not play a role. Furthermore the metric is well approximated as
flat near the conifold point with deviations from flatness leaving the behaviour of $q^{au}$ unaltered\footnote{For a more precise analysis we refer the reader to \cite{Mohaupt:2004pr} where it was shown that the dynamics of the system are 
largely unaffected by the corrections to the moduli space of $Z$.}, and so we may consider 
this action to be valid near the conifold point. The third issue is that
the manifold spanned by the scalars $q^{au}$ is taken to be flat while in a supergravity they would span a quaternionic manifold.
This issue has been addressed in \cite{Lukas:2004du} where it was argued that such a manifold approaches the flat limit for small
values\footnote{By small we mean less than unity in our units.} 
of $q^{au}$ and so within that approximation the flat manifold is valid. This was confirmed 
in \cite{Mohaupt:2004pq,Mohaupt:2004pr} where an example of a quaternionic manifold was compared with the case of a flat manifold
and was found to agree qualitatively. 

\subsection{Moduli trapping}
\label{sec:modulitrapping}

It can be seen from the action that the two types of fields $q$ and $Z$ act as effective masses for each other. In the case
where they are both non-vanishing they will both drive each other to zero. It was shown in \cite{Lukas:2004du} that generically they 
will perform Hubble damped oscillations about the 
conifold point eventually settling at the conifold point $Z=0,q^{au}=0$. 
This is the effect of moduli trapping. 
However, it was also shown that the initial conditions play a very important role in determining the fate of a conifold transition. Since 
we do not have an understanding of what determines the initial conditions, a study of such a system should consider all possible situations and
classify the possible outcomes. This approach leads to outcomes other than the trapping scenario discussed above. 
If the initial value of $Z$ is much larger than that of $q^{au}$ it can happen that the $q^{au}$ settle at zero,
at which point the potential vanishes, before $Z$ is attracted to the conifold point. In that case the conifold point,
$Z=0$ and $q^{au}=0$, is never reached
and the $q^{au}$ are trapped at zero vev but the $Z$ remains a flat direction.
More interestingly if $Z$ begins small and $q^{au}$ begins large\footnote{Throughout this paper by large we mean less than unity but larger than $Z$.} 
it is possible for $Z$ to settle at 
zero while the $q^{au}$ still have reasonably large vevs. In that case
it can be seen from (\ref{action}) that the potential is non-vanishing
 for general values of $q^{au}$ but has flat directions along
\be
 q^{1u}= \pm q^{2u} \label{higgs}
\ee
The $q^{au}$ will therefore be driven towards this point as well as towards zero. If they manage to satisfy (\ref{higgs}) they will
freeze at those values. In that case the Higgs transition is complete and it can be seen from (\ref{action}) that the 
vevs of the $q^{au}$ act as a mass for the gauge field $A^{\mu}$. Geometrically the vev of the $q^{au}$ corresponds to the size
of the two-cycle on the other side of the transition and so if
this is large, and the vev of $Z$ is zero, the transition can be said to have been completed.
The vev of the $q^{au}$ are flat directions and correspond to the modulus associated with the two-cycle and so although $Z$ is trapped
at the conifold point the $q^{au}$ remain as flat directions. It is important therefore to note that, unlike in flop transitions, in conifold 
transitions there are moduli that may not get trapped. 
 
Cosmology, through Hubble friction, plays a crucial role in the processes discussed above. In the case where the transition is completed
 what happens dynamically is that the small $Z$ will oscillate about zero and the $q^{au}$ will perform oscillations about (\ref{higgs}).
To complete the transition both of these oscillations must be damped sufficiently fast so that $Z$ sits at zero and the $q^{au}$ sit at (\ref{higgs}) before 
 the vevs of the $q^{au}$ reach zero. 

Another approach to moduli trapping is through considerations of quantum effects near the conifold point. More specifically it was 
pointed out in \cite{Kofman:2004yc} that as $Z$ oscillates about the conifold point it will produce particles associated with the 
non-perturbative states that are light in the vicinity of the conifold point, these will in turn drive $Z$ further towards the conifold
point thereby trapping it. This is a slightly different scenario to the one discussed above in that 
instead of beginning at an initial configuration of stationary $Z$ and some non-zero vev for $q^{au}$ the idea is to start with zero
vev for $q^{au}$ and an initial velocity for $Z$ in the direction of the conifold point. If both $Z$ and $q^{au}$ are near the conifold
point it is also possible to consider quantum fluctuations creating particle pairs of each type thereby keeping them at the conifold
point even though classically there are flat directions away from the conifold point.
It should be noted though that, as discussed above, in order to complete a conifold 
transition we should have a fairly large vev for $q^{au}$ at the point when $Z$ has settled at the conifold point, and so 
quantum particle production of $Z$ is negligible as the particles associated with $Z$ will be massive.
This means that quantum effects do not change the classical scenario discussed above of completing a transition.

\section{Simulations}
\label{sec:simulations}

In this section we will describe the numerical approach taken to
simulate the model (\ref{sec:num}), and report the results obtained
for different simulations, both for the influence of inhomogeneities
in the cosmological evolution of conifold transitions (\ref{sec:vevs})
and in the possible formation of vortex-like configurations
(\ref{sec:strings}).

\subsection{Numerical details}
\label{sec:num}

In order to perform numerical simulations of the model given by action
(\ref{action}) we use techniques from Hamiltonian lattice gauge
theories
 \cite{Moriarty:1988fx}. 
The usual lattice link and plaquette operators are given by
\ba
& &U_i(x)=e^{-ielA_i(x)}\,;\\
& &Q_{ij}=U_j(x)U_i(x+x_j)U^\dagger_j(x+x_i)U^\dagger_i(x)\,,
\ea
respectively, where $l$ is the lattice spacing, the label $i$ takes the values
$1,2,3$ corresponding to the three spatial dimensions, and $A_i$ are the
gauge fields (the gauge choice $A_0=0$ has been made). 
By $x+x_i$, we denote the nearest lattice point in the $i$ direction from $x$. 
The plaquette operators are related to the gauge field strength  \cite{Moriarty:1988fx},
and the lattice link operator  is used to define
discrete covariant derivatives
\ba
D_i\phi^{1}(x)&=&\frac{1}{l}\left(U_i(x)\phi^{1}(x+x_i)-\phi^{1}(x)\right)
\,;\nonumber\\
D_i\phi^{2}(x)&=&\frac{1}{l}\left(U\dagger_i(x)\phi^{2}(x+x_i)-\phi^{2}(x)\right)\,.
\ea
where $\phi^{1}$ corresponds to both $(q^{11}+iq^{12})$ and $(q^{13}+iq^{14})$; and $\phi^{2}$ to $(q^{21}+iq^{22})$ and $(q^{23}+iq^{24})$. 
Using the lattice link and plaquette operators, we transform  \ref{action}
into a discretized Hamiltonian and derive the discretized equations of motion in the standard
way. 
These equations were solved numerically in a cubic lattice using a staggered leapfrog
method. Several lattice spacings, time steps and cube sizes were
 used in order to check the code, and in fact the results were fairly
 insensitive to these parameters. The actual plots shown in this work
 are for a $200^3$ cube with a ratio of time step to lattice spacing 
 $dt/l=0.2$. We also monitored Gauss's Law throughout the simulations
 to check the stability of the code.

There are two main ingredients in the simulations which are not very
well constrained from the model: initial conditions and the damping
term. We will describe how we treated these two issues in the
remaining of this section, but we would like to point out that a vast
set of simulations was performed to try to cover as many possibilities
as possible, and the main results of the paper remain valid for all
the simulations.

Choosing initial conditions for this cosmological transition is not a
trivial task, since we do not know how the actual transition would
manifest itself. We could have situations in which the field $Z$ and its
velocity are zero, or one where the position or velocity  (or both) are non
zero. The same applies to the scalar fields $q^{au}$. For reasons explained
below, we will focus on the case where $Z$ and its velocity are set to
zero (as mentioned before, we did check that a non-zero value of $Z$
does not change our results).

The consequences of starting with inhomogeneities in the fields
$q^{au}$ or their velocity was investigated, and we realized that the
most effective way of understanding the effect on the transition was by
starting with zero velocity and inhomogeneities in the scalar fields
$q^{au}$. Basically, starting with zero fields but non-zero velocities
resembles the case studied after a few time steps. Furthermore, for
the study of formation of defects, we rely on previous works showing
that the evolution of related systems is fairly insensitive to the
initial condition in the formation of defects 
\cite{Achucarro:1997cx,Urrestilla:2001dd}.
Therefore the initial condition chosen is given by some homogeneous 
value of the scalar fields $q^{au}$, which will be perturbed by some
inhomogeneities, as described below.

As mentioned in section (\ref{sec:action}), one would expect Hubble
friction in the equations of motion coming from the Ricci scalar in
the supergravity theory, and it was shown in
\cite{Lukas:2004du}  that in the homogeneous case the Hubble
damping is an important ingredient in the evolution of the system. We
inherited that result, and included it in our simulation by adding a damping term $\eta$
proportional to the square root of the average energy density of the simulation
in order to mimic Hubble damping.

\subsection{Cosmological evolution}
\label{sec:vevs}

Out of the cases discussed in section \ref{sec:modulitrapping} we will be mainly interested in the possibility of completing 
the transition. The case where $Z$ never reaches the conifold point is rather trivial and the case where both $Z$ and $q^{au}$ are
sitting at the conifold point does not contain any interesting dynamics.

Let us consider the homogeneous case initially. We see that there are three important parameters in the initial conditions of the system
that determine whether we can complete the transition or not. They are the initial vev for $Z$, $\langle Z_0 \rangle_x$, the initial vev for the
$q^{au}$, $\langle q^{au}_0\rangle_x$ (where the subscript $x$ denotes averaging over space) 
 and the initial value for the 'Higgsing' parameter $\Delta$ defined as
\be
\Delta \equiv \sum_{u} \left( \left| \langle q^{1u}_0\rangle_x \right| - \left| \langle q^{2u}_0\rangle_x \right| \right)^2 \label{bigdelta}
\ee  
The parameter $\Delta$ is defined so that it vanishes when the
transition is completed. 
In order to complete a transition we therefore need a configuration of small or vanishing $Z_0$, large $q^{au}_0$ and small $\Delta$. We will not
concern ourselves with the dynamics of $Z$ as these are quite simple and remain unchanged under inhomogeneities. We will therefore set
$Z_0=0$ for the purpose of looking at the possibility of completing a transition whilst keeping in mind that an initial non-zero 
value for $Z_0$ will make the transition harder to complete. 

By introducing spatial inhomogeneities in the values of the fields we introduce a new important parameter $\tilde{\Delta}$, which measures
the effect the spatial inhomogeneities have on the Higgsing parameter. The initial configuration we
chose to simulate are given by a homogeneous vev for the $q^{au}$
given by $\langle q^{au}_0\rangle_x$, and superimposed on that, some
random inhomogeneities:
\be
q^{au}_0(x) = \langle q^{au}_0\rangle_x + \delta \; \hat{n}^{au}(x) 
\ee
where $\delta$ measures the size of the inhomogeneities and the unit vector, $\mathbf{\hat{n}^a}$, randomly distributes 
the inhomogeneities among the hypermultiplet members. 
We can define
\be
\tilde{\Delta} \equiv \delta^2 \sum_{u} \left< \left(  n^{1u}(x) 
 - n^{2u}(x) \right)^2 \right>_x
\ee
where $\langle..\rangle_x$ corresponds to a spatial average. 

We then introduce a total 'Higgsing' parameter, ${\cal D}$, to indicate the effect of inhomogeneities
\be
{\cal D} \equiv \Delta + \tilde{\Delta}
\ee
where $\Delta$ is calculated using the homogeneous part of $q^{au}$.

The perturbations themselves may arise from a number of physical factors: 
quantum fluctuations, finite temperature fluctuations or
inhomogeneous effects in the brane wrapping mechanism. 
For the case where they arise from thermal flunctuations we can get an order of magnitude 
estimate of $\left< \delta \right>^2_T \sim T^2$. 
It is also possible that there will also be inhomogenieties in the brane wrapping mechanism that sets the initial distribution of
$q^{au}$. We do not have an intuition as to the size of these inhomogenieties as this would require detailed knowledge of the 
possible early universe brane gas. Due to these uncertainties, we encode the effects of inhomogeneitis in the parameters $\Delta$ and $\tilde \Delta$, and consider different ranges on their values.

\begin{figure}
\center
\epsfig{file=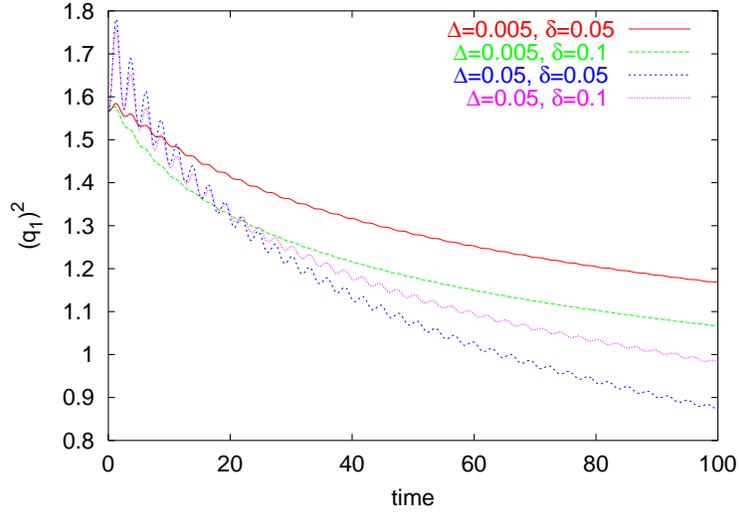,width=10cm}
\flushleft
\caption{
Figure showing the evolution in time of $\left( q^1 \right)^2$ for varying amplitudes of $\Delta$ and $\delta$.}
\label{fig:1}
\end{figure}

Let us consider what kind of effects inhomogeneities will have on the system. We expect there to be an increase in the 
Hubble friction due to the increase in the energy of the system through the contribution of gradient energies. We also expect 
to generate gauge fields due to currents, and finally we see that $\tilde{\Delta}$ will contribute positively to ${\cal D}$. 
The last two effects will make it more difficult to complete the transition, as the gauge fields will have an energy density which is 
minimised at $q^{au}=0$ and so will drive the $q^{au}$ towards zero, and a larger ${\cal D}$ means it takes longer to Higgs. 
The increase in damping however will help complete the transition as was discussed in section \ref{sec:modulitrapping}.
We therefore expect two different regimes to emerge where one effect dominates over the other. The regimes can be parameterised
as 
\ba
\mathrm{Case \;I\; :}&\;& \Delta >> \tilde{\Delta} \\
\mathrm{Case \; II \; :}&\;& \Delta << \tilde{\Delta} 
\ea
Then in case I larger perturbations will help complete the transition, as increasing the fluctuations will
increase the gradient energy and so the Hubble damping, slowing down the fields and enabling them to settle
at a non zero value.
In case II the fluctuations are large to start with, increasing them further will just drive the $q^{au}$
towards zero, i.e. the conifold point. 
We can see this behaviour in figure \ref{fig:1}. The figures show the spatial average of the 
quantity $\left(q^1\right)^2$, defined as
\be
\left(q^1\right)^2 \equiv \frac{1}{V} \sum_u \sum_x \left( q^{1u} \right)^2 \mathrm{,}
\ee
against time for various sizes of $\delta$ and $\Delta$. The vev of $\left(q^2\right)^2$ followed the same type of evolution as $\left(q^1\right)^2$ with
both oscillating about each other. Figure \ref{fig:1} shows how the possibility of completing the transition is manifested in the field theory. The vevs of 
$\left(q^{au}\right)^2$ tend towards a non-zero asymptotic value that corresponds to the size of the two-cycle on the other side. The magnitude of the asymptotic
value therefore determines whether the transition is completed or the moduli are trapped.
The two lines with $\Delta \sim \delta$ ($\Delta=0.05$) correspond to case I and we see that increasing $\delta$ 
increases the asymptotic value for $\left(q^1\right)^2$ thereby helping complete the transition with a large two-cycle on the other side. 
The two lines with $\Delta << \delta $ ($\Delta=0.005$) correspond to case II and here we see that larger $\delta$ drives 
the asymptotic value further towards zero and so a small size for 
the cycle. 

Having shown that there are two regimes with quite different behaviour we might speculate on which is the more 
physical.
The first regime is  when $\Delta >> \tilde{\Delta}$ (case I).
Physically this situation corresponds to 
the case where the spatial averages of the number of branes wrapping each of
the two cycles differ substantially and the spatial perturbations of the number of wrapping branes 
are small in comparison to this difference. It is difficult to think
of a scenario in string theory where such an initial condition could come about. The reason for this 
is that the two cycles are homologically related and so both must 
degenerate simultaneously. Given that they have an equal size the masses of states wrapping them are 
also equal and so it is unlikely that there will be much more
of one than the other. The second regime occurs when $\Delta << \tilde{\Delta}$ (case II). Physically this 
scenario corresponds to the case where the difference in the number of 
wrapping branes arises primarily due to spatial perturbations. This is a more likely
scenario and in fact should be the case generically as it 
simply corresponds to the finite correlation length of the system. In this case the inhomogeneities 
help to trap the remaining moduli thereby stopping the transition from completing.

\begin{figure}
\center
\epsfig{file=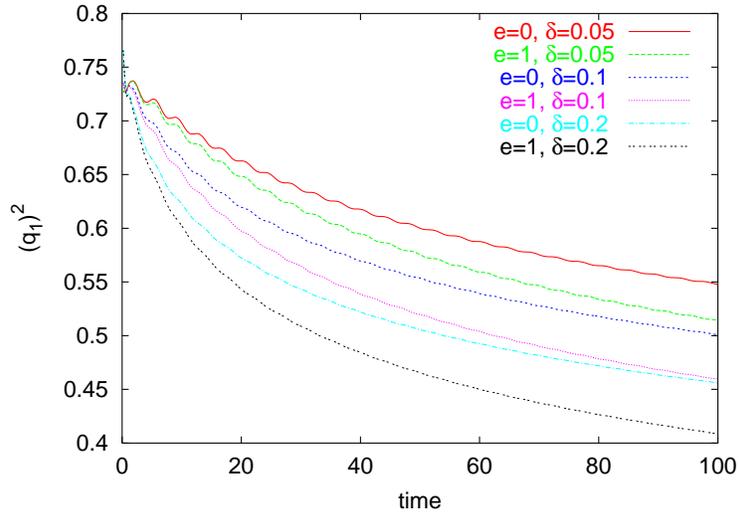,width=10cm}
\flushleft
\caption{
Figure showing the effects of coupling to gauge fields on the evolution of $\left( q^1 \right)^2$.}
\label{fig:2}
\end{figure}

The effect of gauge fields is shown in Figure \ref{fig:2}. We see plots for various perturbation sizes 
with the charge of the hypermultiplet fields, $e$, on and off.
We see that coupling the hypermultiplet to the gauge fields that are naturally induced always drives their vev towards zero thereby helping to trap them and 
hindering the completion of the transition. The plots shown are for the case where the initial conditions are of no gauge fields and so the gauge fields present are
the ones induced through the currents generated since the beginning of the simulation. There is of course the possibility of some initial gauge field density and this
will amplify the effects shown in the simulations. 

\subsection{String formation}
\label{sec:strings}

In \cite{Greene:1996dh} it was suggested that flux-tube solutions to
the potential in (\ref{action}) could form, due to the existence of a
quantized magnetic flux. However, 
in \cite{Achucarro:1998er} it was shown that those 
infinite axi-symmetric
flux tubes solutions 
 are unstable to the expansion of the core.

In ordinary Abrikosov-Neilesen-Olensen vortices (see for instance
\cite{vilenkin:1994,Hindmarsh:1994re}),
topology ensures that there will be lines of zeros of the scalar field, and
those zeros will correspond to lumps of potential energy. The gauge
fields are massless around those zeros, and the magnetic flux-tubes
will follow the lines of zeros. The core of those defects is
determined by the competition between the magnetic field lines wanting to
spread apart and have a thick core, and the potential energy wanting
to have a small core.

When global symmetries are present in the system (like in the one we
are investigating), the situation can be more complicated. In, for
example, the semilocal model \cite{Achucarro:1999it},  topology does not
 ensure
that there will be any line of zeros of scalar fields. The question of
the existence and stability of the strings is purely dynamical and
depends on the parameters of the system. One still has that a zero of
the scalar field necessarily means some concentration of potential
energy; and the competition of magnetic field versus potential energy
is still there. However, the system has the possibility of not forming
a core by gaining gradient energy instead. Depending on the
parameters, we can have stable, unstable or neutrally stable strings.

In the model given by (\ref{action}), the situation is similar to the
semilocal case, but here the existence of a zero in the scalar fields
does not mean that there will be a concentration of potential
energy. So topology does not ensure the existence of zeros, and there
is no competition between magnetic field and potential energy. So, as
shown in \cite{Achucarro:1998er}, a string in this system will tend to broaden its
core, and is unstable.

Therefore, it was speculated that the flux density could form
spherical shells. The actual existence and stability of these shells is
unknown, though the arguments given above would suggest that the
magnetic flux would dilute fast, as there is no mechanism holding it together.
We performed many simulations of this field theory, using a variety of
initial conditions. We expected some magnetic flux to be formed, seeded
by the inhomogeneities, which would act as mass terms for the scalar
fields $q^{au}$ and would make them evolve to zero, maybe forming
some structure.
Observing level surfaces of the magnetic energy density  we aimed at
looking for such structure. However, even though some magnetic field was excited,
we did not observe any shell. What we
found was that the scalar field would form lumps while the gauge field 
flux appeared simply as a white noise background, not following the scalar field. 
This could simply
be a result of not being able to find a configuration in which the induced
magnetic field lived long enough so as to be able to follow the
scalar field, and have enough time to form some structure. In all our
simulations, the magnetic field decayed (the flux diluted very fast). Bear in
mind that the damping is quite high in the initial states of the
simulation, delaying the gauge field from tracking the scalars. In any case, our
simulations give very strong evidence that
there are simply no vestiges of the unstable vortices.

\section{Conclusions}
\label{sec:conclusions}

In this paper we considered the effects that inhomogeneities in the fields have on the cosmology 
of type IIB conifold transitions. By performing three dimensional
simulations of the transition we showed that inhomogeneities in the fields can either help or 
hinder the completion of a transition. 
The physics which would imprint the inhomogeneities includes
quantum fluctuations, thermal fluctuations and the details of the brane wrapping mechanism.
Due to this uncertainty we have
outlined two regimes, parameterised
by the relative magnitude of the inhomogeneities ($\tilde\Delta$) and the 'Higgsing' parameter $\Delta$,
to get a physical picture of the dynamics.
The first regime is when 
local fluctuations are small ($\Delta >> \tilde{\Delta}$) and
in this case the inhomogeneities help the transition complete. 
The second case is where the local
fluctuations dominate, $\Delta <<\tilde\Delta$. Here we find that increasing the inhomogeneity
tends to trap the moduli at the conifold point.

A second issue in this model is that of vortices,
if vortices could actually form in this
 system, it would have been been interesting since it might have
 been a possibility of some experimental signature of conifold
 transitions.
 It was already known
 that these vortices were unstable, but we were 
able to use our simulations to check for any effect the
 unstable vortices may have, in case they were formed and lasted long enough. 
We did not find any trace of the unstable vortices, and moreover, the
speculated spherical balls of flux that could be formed instead were
 also not observed.
This gives strong
evidence that there is no structure formed by the magnetic field. 

\section*{Acknowledgements}

The authors would like to thank Ana Ach\'ucarro and Mark Hindmarsh for useful
discussions. E.P and P.S. are funded by PPARC. J.U. is supported by
the spanish {\it Secretar\'{\i}a de Estado de Educaci\'on y
  Universidades} and {\it Fondo Social Europeo}. This work was
partially supported by FPA 2002-02037, 9/UPV00172.310-14497/2002 and
 FPA2005-04823. We also acknowledge the extensive use of the UK National
Cosmology Supercomputer funded by PPARC, HEFCE and Silicon Graphics.


\end{document}